\documentclass[prb,amsmath,amssymb,onecoloumn,longbibliography,superscriptaddress]{revtex4-2}

\usepackage{amsmath}
\usepackage{amssymb}
\usepackage{graphicx}
\usepackage{siunitx}
\usepackage{hyperref}

\begin{document}

\title{Coulomb blockade thermometry beyond the universal regime}

\author{Nikolai Yurttagül}

\affiliation{QuTech and Kavli Institute of Nanoscience, Delft University of
Technology, 2600 GA Delft, The Netherlands.}

\affiliation{VTT Technical Research Centre of Finland Ltd, P.O.~Box 1000, FI-02044 VTT Espoo, Finland}

\author{Matthew Sarsby}

\affiliation{QuTech and Kavli Institute of Nanoscience, Delft University of
Technology, 2600 GA Delft, The Netherlands.}

\author{Attila Geresdi}

\email[e-mail: ]{geresdi@chalmers.se}

\affiliation{QuTech and Kavli Institute of Nanoscience, Delft University of
Technology, 2600 GA Delft, The Netherlands.}

\affiliation{Quantum Device Physics Laboratory, Department of Microtechnology and Nanoscience, Chalmers University of Technology, SE 41296 Gothenburg, Sweden}

\begin{abstract}
The charge localization of single electrons on mesoscopic metallic islands leads
to a suppression of the electrical current, known as the Coulomb blockade. When
this correction is small, it enables primary electron thermometry, as it was
first demonstrated by Pekola et al. (Phys. Rev. Letters, 73, 2903 [1994]).
However, in the low temperature limit, random charge offsets influence the
conductance and limit the universal behavior of a single metallic island. In
this work, we numerically investigate the conductance of a junction array, and
demonstrate the extension of the primary regime for large arrays, even when the
variations in the device parameters are taken into account. We find that our
simulations agree well with measured conductance traces in the submillikelvin
electron temperature regime.
\end{abstract}

\maketitle

\section{Introduction}

Single electron charging effects are prominent in isolated nanoscale conductors
below a characteristic temperature scale $k_\textrm{B} T \ll e^2/2C_\Sigma$
\cite{averin_inbook_1991,singlechargetunneling_book}, where $k_\textrm{B}$ is the
Boltzmann constant, $e$ is the electron charge and $C_\Sigma$ is the total
capacitance of the metallic island. In this low temperature regime, charge sensing
\cite{PhysRevLett.70.1311} and direct current measurements through tunneling contacts
with low conductance, $G \ll e^2/h$ \cite{PhysRevLett.59.109} yield a
well-defined number of excess electrons on small metallic islands and an
exponentially suppressed tunneling current.
Increasing the temperature to $k_\textrm{B} T > e^2/2C_\Sigma$ results in the breakdown
of Coulomb blockade, and it was shown that in this regime, the conductance
suppression is universal. This enables a primary measurement of the electron
temperature \cite{Pekola1994}, which is insensitive to the device geometry, the
external magnetic field \cite{Pekola1998} and the electrostatic environment of
the device \cite{hahtela2013investigation}.
This property makes Coulomb blockade thermometry favorable for metrological
applications \cite{doi:10.1063/1.371296,doi:10.1063/1.1351526,Meschke2011,Hahtela2017}, and several experiments were
performed in cryogenic temperatures ranging down to the millikelvin
\cite{KNUUTTILA1998224,PhysRevLett.101.206801,Casparis2012,Bradley2016,Palma2017,Bradley2017,Yurttaguel2019} and submillikelvin regime
\cite{Sarsby2020} as well as up to $60\,$K \cite{Meschke2016}.

The practical upper temperature range of a Coulomb blockade thermometer (CBT) is
limited by the readout of the low bias conductance suppression, which is
inversely proportional to the temperature \cite{Meschke2011,Meschke2016}. In
the low temperature limit, the conductance depends on the electrostatic
environment of the device, which imposes an effective offset charge
\cite{PhysRevLett.59.109,GEERLIGS1990973}.
This offset charge depends on the random, uncontrolled population of charge
traps at nearby interfaces and charged impurities inside dielectrics
\cite{doi:10.1063/1.108195,PhysRevB.56.7675} leading to a statistical error
for primary electron thermometry by CBTs.

Precision electron thermometry requires a characterization of this error source.
In particular, the recent advent of ultralow-temperature quantum electronics
\cite{RevModPhys.78.217,pickett2018,Jones2020} requires electron
thermometry spanning several orders of magnitude in temperature. Here, we use a
statistical Monte Carlo approach to investigate the temperature error sources of
CBTs in the low temperature regime, and demonstrate that this numerical
procedure provides a unified description of a CBT, matching the predictions of
the commonly used single island master equation (ME) model \cite{Pekola1994} in
the high temperature, universal regime. Importantly for Coulomb blockade thermometry, we
show that the effect of random offset charges is suppressed in the case of large
arrays, which enables reliable temperature measurements deeper in the low
temperature regime, even in the presence of variations in the device parameters.
Finally, we demonstrate the applicability of our numerical results by a direct
comparison with ultralow-temperature experimental data, where previous numerical
and analytical models fail to describe the conductance of the CBT.

\section{Electrostatics of the CBT device}

\begin{figure}
\centering
\includegraphics[width=\textwidth]{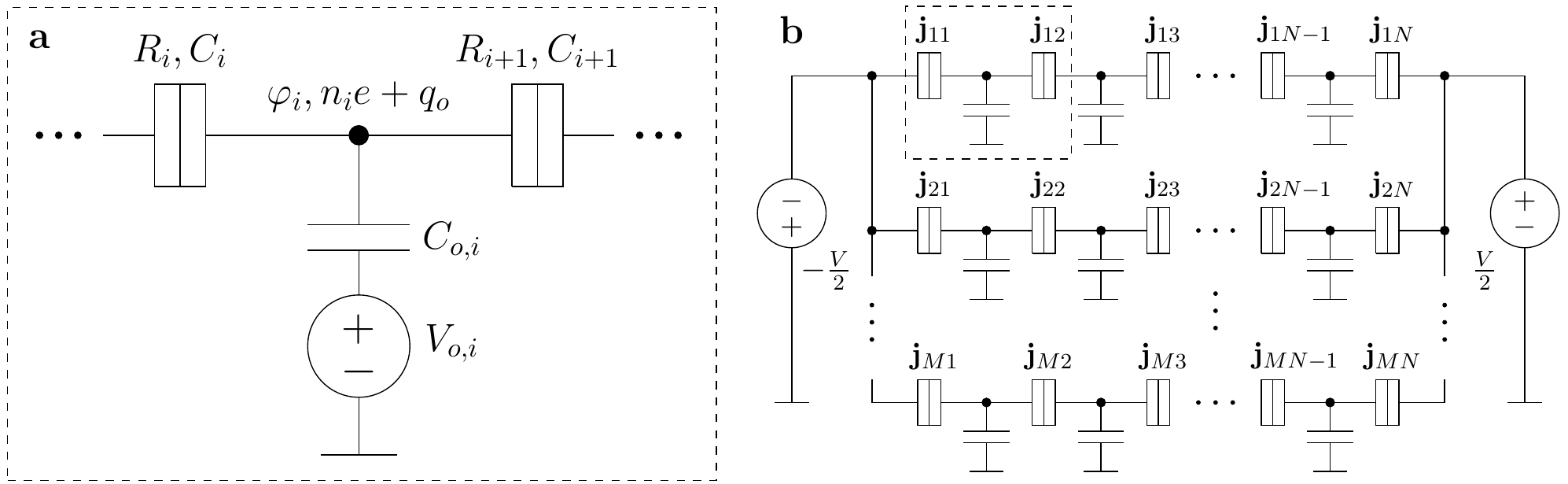}
\caption{The schematics of the tunnel junction array. (a) A single island with
an offset charge of $q_{\textrm{o},i}=-C_{\textrm{o},i}V_{\textrm{o},i}$, where $C_{\textrm{o},i} \ll C_i, C_{i+1}$. (b) The array
with $N$ junctions in series and $M$ rows in parallel subject to a symmetric voltage bias
of $V$. A single island depicted in (a) is enclosed in the dashed box.}
\end{figure}

We consider the device geometry outlined in Fig.~1, which displays the typical
implementation of a CBT consisting of $N$ tunnel junctions in series enclosing
$N-1$ islands. Each junction possesses a resistance $R_i$ and a parallel
capacitance $C_i$. Crucially for the current work, we also consider the
offset-charge of each island,
$q_{\textrm{o},i}=-C_{\textrm{o},i}V_{\textrm{o},i}$ yielding a total charge
$q_i=en_i + q_{\textrm{o},i}$, where the number of excess electrons on the
island is $n_i$. Throughout this work, we assume that the effective gating
capacitance $C_{\textrm{o},i}$ is much smaller than $C_i$ and $C_{i+1}$ for all
islands and investigate the applicability of this limit for experiments in Section 7.

The array is symmetrically biased by two voltage sources, fixing the
electrostatic potential on the leftmost and rightmost nodes at $-V/2$ and $V/2$
respectively, leading to a total voltage drop of $V$. We neglect
environment-assisted tunneling processes \cite{Ingold1992} assuming a low-impedance
electromagnetic environment surrounding the device.

In our analytical and numerical calculations, we always consider a single series
of junctions first and sum up the resulting conductances to model $N\times M$
arrays. This method is equivalent to neglecting the capacitive coupling between
islands in neighboring chains, which is justified in typical device geometries.

We now consider the electrostatic potential $\varphi_i$ of each island for
$i=1\ldots N-1$, which allows us to express the total charge on the island:
\begin{equation}
C_i\varphi_{i-1}-\left(C_i+C_{i+1}\right)\varphi_i+C_{i+1}\varphi_{i+1}=-q_i,
\end{equation}
where $\varphi_0=-V/2$ and $\varphi_N=V/2$.
We can write this equation in a matrix form, $\mathbf{C}\cdot \mathbf{\varphi} = - \mathbf{q}$ with 
\begin{align}
\mathbf{C}=
\begin{pmatrix}
-C_{\Sigma 1}&C_2 & & &\\
C_2&-C_{\Sigma 2}&C_3 & &\\
& \ddots & \ddots &\ddots& \\
 & C_{N-2}& -C_{\Sigma (N-2)} & C_{N-1}\\
& & C_{N-1} & -C_{\Sigma (N-1)}\\
\end{pmatrix},
\end{align}
where $C_{\Sigma i}=C_i + C_{i+1}$, the total capacitance of island $i$.

Finally, the electrostatic potentials define the voltage drop over each tunnel
junction as $V_i=\varphi_{i+1}-\varphi_i$ and the total electrostatic energy is
given by
\begin{equation}
F=\frac{1}{2}\sum_i C_i V_i^2.
\end{equation}

We now consider a homogeneous array in the absence of offset charge, where all $C_i=C$ and $q_{\textrm{o},i}=q_\textrm{o}=0$,
at zero bias voltage, $V=0$. In this limit, the electrostatic energy of a single
electron on island $k$ is calculated using the net capacitance $C(k^{-1}+(N-k)^{-1})$, which yields
\begin{equation}
F=\frac{e^2}{2C}\frac{k(N-k)}{N}.
\end{equation}
Substituting $C_\Sigma=2C$ and $k=1$ or $k=N-1$, Eq.~(4) yields the charging energy expression
\begin{equation}
E_\textrm{C}=\frac{e^2}{C_\Sigma}\frac{N-1}{N},
\end{equation}
which defines the dimensionless inverted temperature $u=2E_\textrm{C}/k_\textrm{B} T$
\cite{Pekola1994,doi:10.1063/1.115090}.

\begin{figure}
\centering
\includegraphics[width=\textwidth]{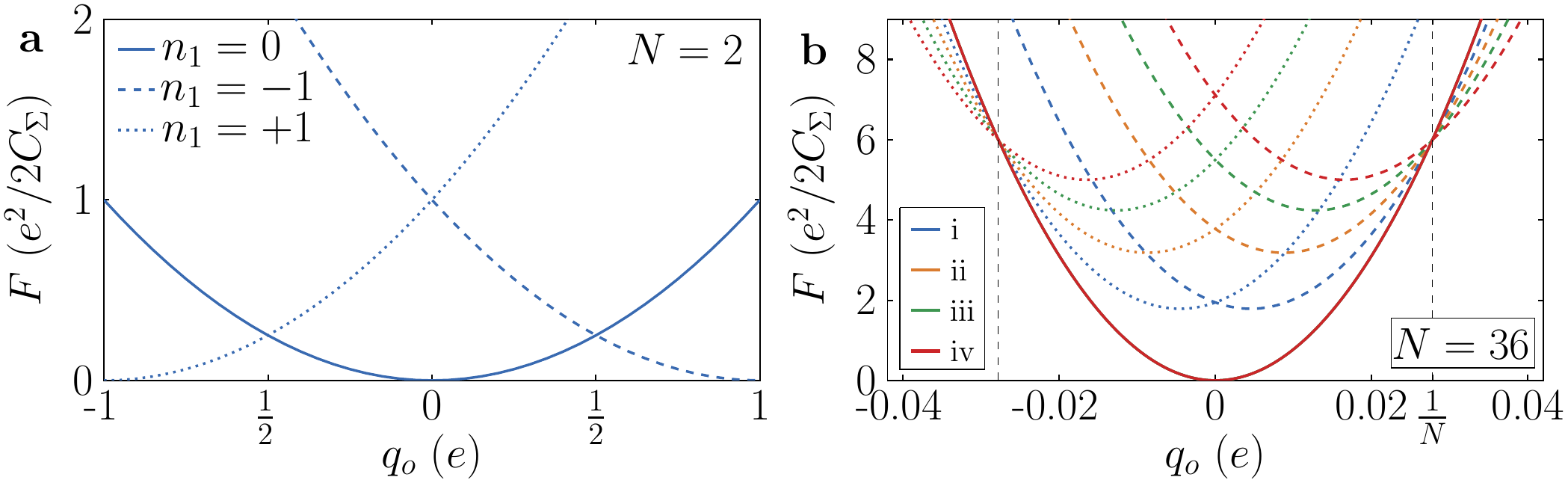}
\caption{CBT electrostatic energy as a function of the offset charge $q_\textrm{o}$. (a)
The single island case ($N=2$) exhibits degeneracy points at half-integer
$q_\textrm{o}/e$, where the island charge $n_1$ can change with 1. (b) For an arbitrary
$N$, the zero excess charge case $\mathbf{q}=\{0,0,\ldots,0\}$ (solid red line)
is degenerate with a single excess electron anywhere on the chain at $q_\textrm{o}/e=-1/N$.
The dashed and dotted lines denote the following configurations, respectively:
i:~$\mathbf{q}=\{\pm e,0,0,0,...,0\}$, ii:~$\mathbf{q}=\{0,\pm e,0,0,...,0\}$,
iii:~$\mathbf{q}=\{0,0,\pm e,0,...,0\}$, iv:~$\mathbf{q}=\{0,0,0,\pm e,0...,0\}$}
\end{figure}

Turning to the effect of a finite offset charge $q_{\textrm{o},i}=q_\textrm{o}$ on all
islands of a homogeneous array, we derive the condition where the change of
$F$ (see Eq.~(3)) is zero in response to a single electron tunneling event
(isoenergetic tunneling). With zero excess electrons on all islands, charge
conservation yields the following expression for the voltage drops over the
tunnel junctions:
\begin{equation}
V_{i+1}-V_i=-q_\textrm{o}/C,
\end{equation}
and $\sum_i V_i=V$. If we now insert a single excess electron onto island
$k$, $V'_{k+1}-V'_k=-(q_\textrm{o}-e)/C$ holds with the other differences remaining the
same. We can now zero out $\Delta F \propto \sum_i V_i'^2 - \sum_i V_i^2$ with
the following index shift:
\begin{align}
V_1' \ldots V_k' &\rightarrow V_{N-k} \ldots V_N \nonumber \\
V_{k+1}' \ldots V_N' &\rightarrow V_1 \ldots V_{N-k-1}
\end{align}
which is fulfilled for any $k$ when $q_\textrm{o}=e/N$, yielding $\Delta F=0$ independent of the bias voltage $V=\sum_i V_i$.

This result has important consequences to the charge degeneracy of the device. The
well-known charge degeneracy condition of $q_\textrm{o}=e/2$ in case of a single island
changes to $q_\textrm{o}=e/N$ for an array of $N$ tunnel junctions, where the total
electrostatic energy of an empty chain is equal to the energy when a single excess
electron is present on any of the $N-1$ islands. We illustrate this
effect by plotting $F$ at $V=0$ as a function of $q_\textrm{o}$ in the case of
$0$ and $\pm e$ excess charge in the case of $N=2$ (Fig.~2a) and $N=36$
(Fig.~2b), where different positions of the excess charge $k=1\ldots 4$ are
shown, all intersecting at $q_\textrm{o}=e/N$. This result shows that in the low
temperature regime, we should expect the highest conductance of a CBT array at
this offset charge corresponding to minimal Coulomb blockade \cite{Feshchenko2013}.

\section{Numerical model}

We now introduce the numerical method leading to the voltage-dependent
differential conductance $G(V)$ of a CBT array at an arbitrary temperature. We base our calculations on the
Markov chain Monte Carlo (MCMC) method \cite{Bakhvalov1988,Wasshuber2001}, which
allows us to investigate arbitrary array sizes, offset-charge
configurations, and variations in the parameters of the tunnel junctions. 

Here the system is described by a chain
of randomly selected single charge tunneling events altering $\mathbf{q}$ to
$\mathbf{q}'$, which is written in a matrix form for $\mathbf{q}^\text{T}=[q_1,q_2,\ldots,q_{N-1}]$:
\begin{equation}
\mathbf{Q'}=
\begin{pmatrix}
q_1-e&q_1+e&q_1&\cdots&q_1&q_1+e&q_1-e&\cdots&q_1\\
q_2&q_2-e&q_2+e&\cdots&q_2&q_2&q_2+e&\cdots&q_2\\
q_3&q_3&q_3-e&\cdots&q_3&q_3&q_3&\cdots&q_3\\
\vdots&\vdots&\vdots&\ddots&\vdots&\vdots&\vdots&\ddots&\vdots\\
q_{N-1}&q_{N-1}&q_{N-1}&\cdots&q_{N-1}+e&q_{N-1}&q_{N-1}&\cdots&q_{N-1}+e
\end{pmatrix}.
\end{equation}
The first (last) $N-1$ columns describe the forward (backward) tunneling of
one charge quantum. Next we evaluate the resulting potentials in a concise form
$\varphi'=\mathbf{C}^{-1}\mathbf{Q'}$ and calculate the associated free energy
change for a single electron hopping through junction $i$ \cite{Bakhvalov1988}:
\begin{equation}
\Delta F_i^{\pm}=\frac{1}{2}\sum_{k=1}^{N}C_k\left(\varphi'_{k}-\varphi'_{k-1}\right)^2-\frac{1}{2}\sum_{k=1}^{N}C_{k}\left(\varphi_{k}-\varphi_{k-1}\right)^2.
\end{equation}
Here, $\varphi'_0=\varphi_0=-V/2$ and $\varphi'_N=\varphi_N=V/2$ at a bias voltage $V$ (Fig.~1b).
We note that for a homogeneous array with $C_k=C$, the above expressions lead to
$\Delta F_i=e/2(\varphi_{i\pm1}'+\varphi_{i\pm1}-\varphi_{i}'-\varphi_{i})$,
which was also used to find the analytic high temperature $G(V)$ curve for the $N=2$ case \cite{Pekola1994}. Based
on $\Delta F_i$, we calculate the forward and backward tunneling rates
through junction $i$ with a resistance $R_i$:
\begin{equation}
\Gamma_i^\pm=\frac{1}{e^2R_i}\frac{\Delta F_i^{\pm}}{1-\textrm{exp}(-\Delta F_i^{\pm}/k_\textrm{B}T)}.
\end{equation}
The tunneling current $I=\pm e\Gamma_i^\pm$ is a result of the single charge tunneling
events. Following standard Monte Carlo methods, we consider the current charge
configuration $\mathbf{q}$ and evaluate the corresponding $\Gamma_i^\pm$ values.
Then, we randomly select the realized tunneling process with a probability
proportional to the corresponding tunneling rate. In addition, we follow the
variance reduction method introduced by \cite{Wasshuber2001,Hirvi1996} to
calculate $I$. Here, we make use of the current conservation $I=I_1=I_2=...=I_N$
along the tunnel junction chain, and express the total current as $I=\Delta
Q/\Delta t$ using \cite{Hirvi1996}:
\begin{equation}
\Delta t=\sum_{p=1}^P\Delta t_p=\sum_{p=1}^P\left[\sum_{i=1}^N\left(\frac{1}{\Gamma_i^++\Gamma_i^-}\right)\right]_p,
\end{equation}
\begin{equation}
\Delta Q  = \sum_{p=1}^P\Delta Q_p=\sum_{p=1}^P\left[e\displaystyle\frac{\sum_{i}\left(\Gamma_i^+-\Gamma_i^-\right)R_i/R_\Sigma}{\sum_{i}\left(\Gamma_i^++\Gamma_i^-\right)}\right]_p,
\end{equation}
where the Monte Carlo step $p$ runs in the range of $1\ldots P$, which we set to
reach a relative error less than $10^{-4}$, typically achieved with $P<10^5$. We
numerically acquire the differential conductance as $G(V)=(I(V+\delta
V)-I(V-\delta V))/2\delta V$, where $\delta V \ll k_\textrm{B} T/e$. Finally, we recover
the asymptotic tunnel conductance at high bias voltages as
$G_\textrm{t}^{-1}=R_\Sigma=\sum_i R_i$.

\begin{figure}
\centering
\includegraphics[width=\textwidth]{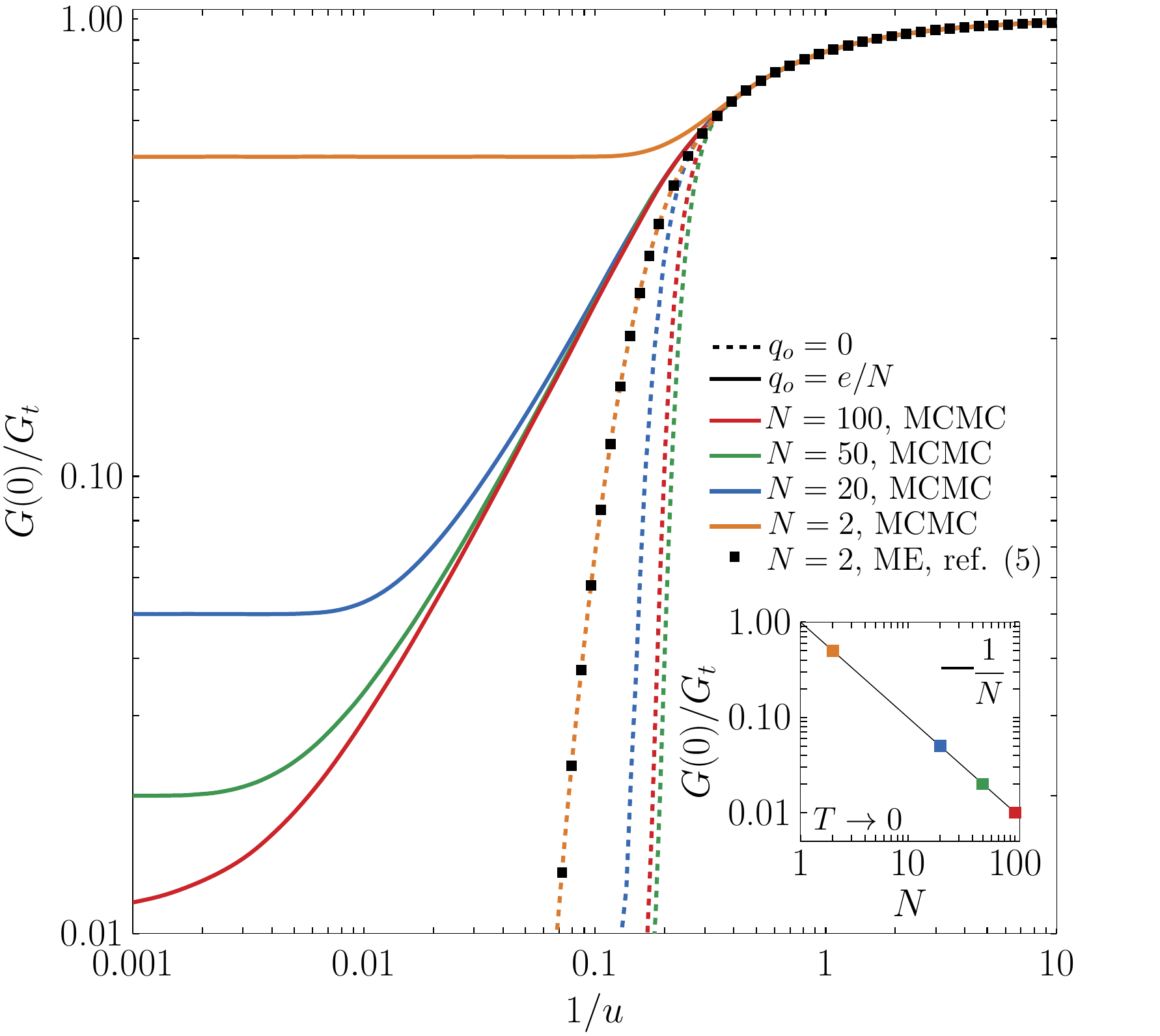}
\caption{The limiting cases for the normalized zero bias conductance $G(0)/G_\textrm{t}$
as a function of the dimensionless temperature $1/u$ for different array
lengths, $N$. The dashed lines show the case of full Coulomb blockade,
$q_{\textrm{o},i}=0$, whereas solid lines are evaluated at the charge degeneracy points,
$q_{\textrm{o},i}=e/N$. The inset displays the calculated conductance in the low
temperature limit for the latter case, and the solid line depicts $G(0)/G_\textrm{t}=1/N$, see Eq.~(15). The black square symbols in the main panel depict the master equation solution
for $N=2$, $q_\textrm{o}=0$ \cite{Pekola1994}.}
\end{figure}

First, we discuss the dimensionless zero bias conductance $G(0)/G_\textrm{t}$, which
we plot in Fig.~3 for various $N$ values and for the two limiting cases in
homogeneous offset charge, $q_\textrm{o}=0$ (dashed lines) and $q_\textrm{o}=e/N$ (solid lines).
We benchmark our numerical MCMC results in the well-studied high temperature
regime, where $u \lesssim 1$. Here, we recover the earlier results
\cite{Pekola1994,Feshchenko2013} that the conductance collapses onto a single
curve independent of $N$ and $q_\textrm{o}$. The conductance in this limit has been
calculated earlier by writing a master equation for forward and backward
tunneling processes for a single island device with $N=2$. Here, the conductance
suppression $\Delta G=G_\textrm{t}-G(0)$ follows a series expansion in $u \ll 1$ \cite{Pekola1994,Feshchenko2013}:
\begin{equation}
\Delta G/G_\textrm{t} = u/6 - u^2/60 + u^3/630 - \ldots
\end{equation}

In the low temperature limit, we find that the conductance depends on $q_\textrm{o}$, and the $q_\textrm{o}=0$
case corresponding to strong Coulomb blockade results in an exponential suppression of
the conductance. However, at minimal blockade, $q_\textrm{o}=e/N$, our results yield a low
temperature saturation conductance corresponding to $G(0)/G_\textrm{t}=1/N$ (see inset of
Fig.~3), which is the extension of the $N=2$ case discussed earlier \cite{Feshchenko2013}.

\begin{figure}
\centering
\includegraphics[width=\textwidth]{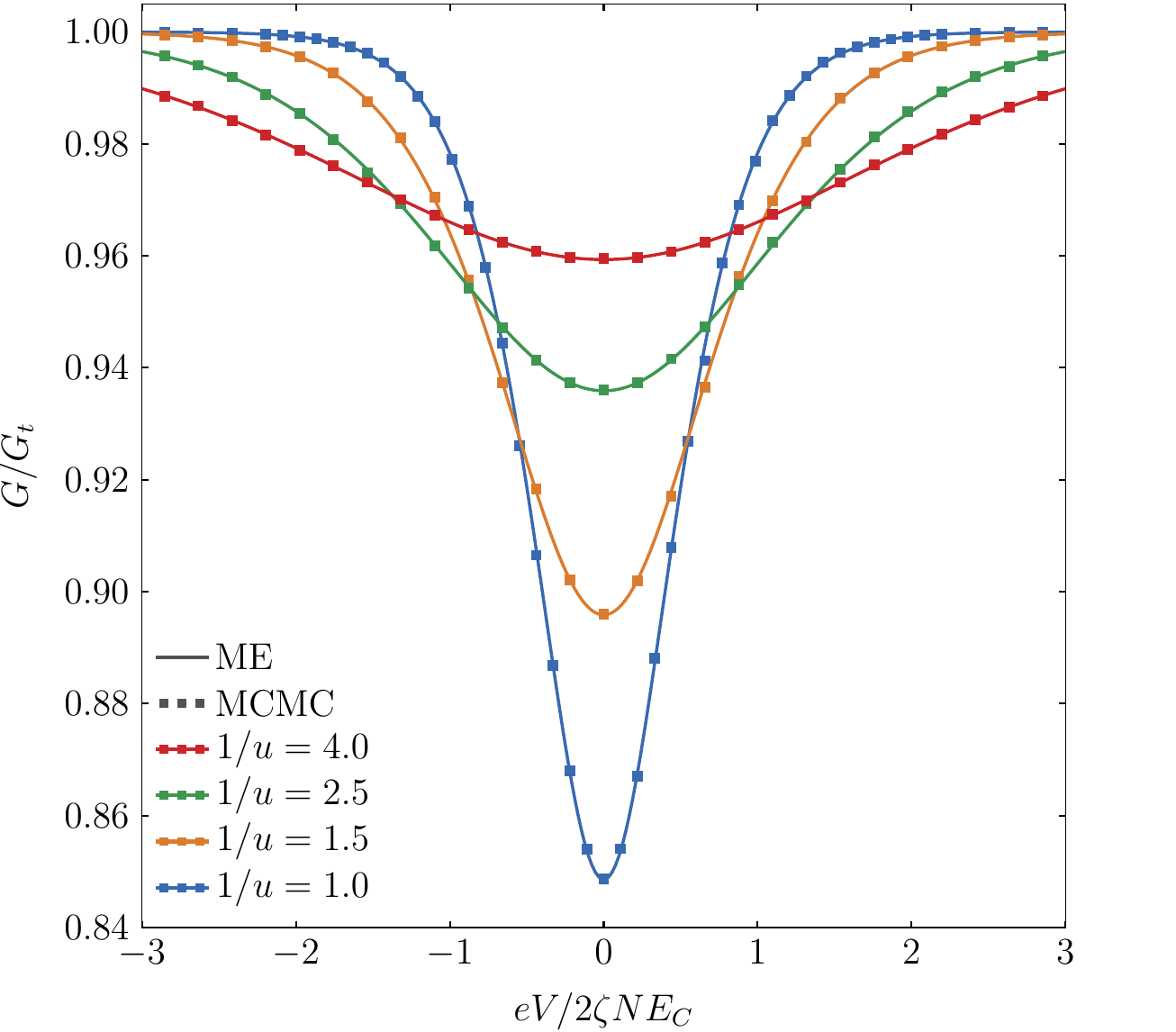}
\caption{The comparison of the master equation (ME) model for $N=2$ and the
Markov chain Monte Carlo (MCMC) simulation for $N=36$ in the high temperature
regime, where $u \le 1$. The normalized differential conductance curves are
displayed as the function of the dimensionless bias voltage, $eV/(2\zeta N E_\textrm{C})$, where $\zeta\approx 5.439$, see Eq.~(16) \cite{Pekola1994}.
The solid lines denote the ME results, and the corresponding MCMC results are
displayed as full squares of the same color.}
\end{figure}

\begin{figure}
\centering
\includegraphics[width=0.75\textwidth]{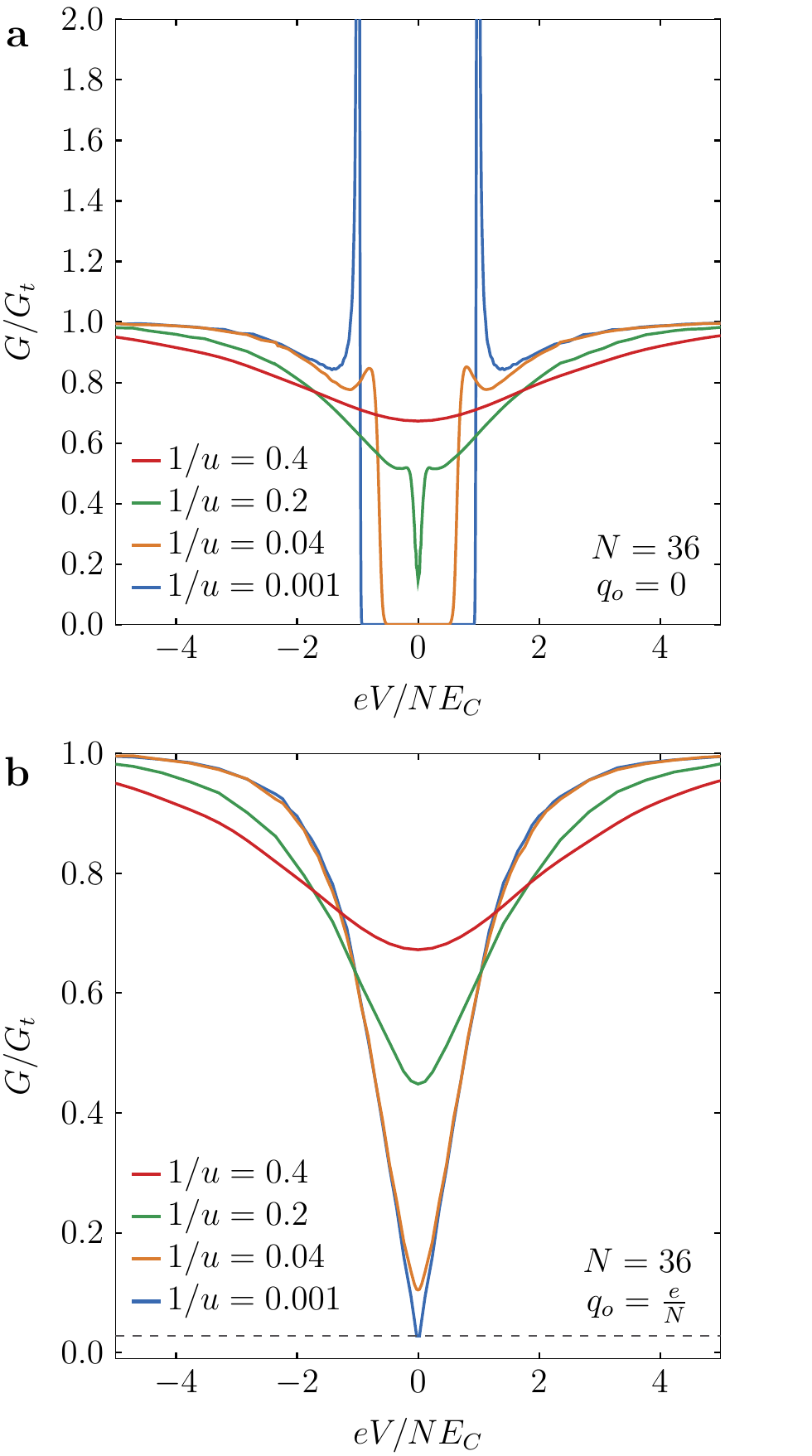}
\caption{The normalized differential conductance curves of a $N=36$ devices as
a function of the dimensionless bias voltage $eV/N E_\textrm{C}$ in the low temperature regime
where $u > 1$, as calculated by the MCMC model. Panel (a) shows the
zero offset charge result, whereas panel (b) displays the minimal blockade curves
with $q_\textrm{o}=e/N$. Here, the horizontal dashed line at $G/G_t=1/36$ displays the analytical zero temperature and zero bias limit Eq.~(15).}
\end{figure}

We understand this result based on the charging energy expression summarized
in Fig.~2, where all $N-1$ possible states with a single excess electron and the
empty chain have the same electrostatic energy if $q_\textrm{o}=1/N$. Since all other
charge configurations are higher in energy, only these $N$ states are occupied
at zero temperature, with an equal probability of $\mathcal{P} = 1/N$. Taking the
$T\rightarrow 0$ limit of Eq.~(10), we find
\begin{equation}
\Gamma=\frac{\Delta F}{e^2R},
\end{equation}
if $\Delta F >0$ (forward tunneling) and $\Gamma=0$ otherwise. We calculate
the current as $I=\mathcal{P}\Gamma$ through any of the tunnel junctions. At low bias, where
$eV\ll E_\textrm{C}$, we get $\Delta F=eV/N$, which yields $I=V/(RN^2)$. With $G=dI/dV$ and $G_\textrm{t}=1/(NR)$, we arrive to the dimensionless conductance 
\begin{equation}
\left(\frac{G(0)}{G_\textrm{t}}\right)_{T\rightarrow 0}=\frac{1}{N}
\label{gt0}
\end{equation}
corresponding to the numerically acquired result. This conductance and its
exponentially suppressed counterpart at $q_\textrm{o}=0$ demonstrates the increasing
sensitivity of a CBT to the offset charge as the temperature is lowered below
$E_\textrm{C}$. Furthermore, in this regime, the commonly used analytical master equation
model with $N=2$ fails to predict $G(0)/G_\textrm{t}$ regardless of $q_\textrm{o}$, underlining the
importance of MCMC numerics for low temperature Coulomb blockade thermometry. 

Next, we perform the analysis of the finite bias $G(V)/G_\textrm{t}$ curves, which is
commonly used for primary calibration in the universal regime. Here, the full
width at half minimum \cite{Pekola1994,doi:10.1063/1.115090}
\begin{equation}
eV_{1/2}= \zeta Nk_\text{B}T, 
\end{equation}
with $\zeta \approx5.439$ enables primary thermometry if $\Delta G/G_\textrm{t} \ll 1$ or the characterization of
$E_\textrm{C}$ for secondary thermometry with $G(0)/G_\textrm{t}$ using Eq.~(13).
We first demonstrate that MCMC calculations reproduce the master equation conductance traces
in the universal regime and yield the same $V_{1/2}$ (Fig.~4).

In the low temperature regime, we observe that $G(V)$ depends on the offset
charge $q_\textrm{o}$ (Fig.~5). The $q_\textrm{o}=0$ case displayed in Fig.~5a was discussed
earlier in the context of charge soliton propagation on tunnel junction chains,
where the threshold voltage was found to be $eV_\textrm{t}= N E_\textrm{C}$ at $1/u
\rightarrow 0$ \cite{Amman1989,PhysRevB.49.16773}, in agreement with our
numerical results. In contrast, the $q_\textrm{o}=e/N$ case shown in Fig.~5b exhibits a
gapless behavior, with no well-defined threshold voltage. However, in agreement
with the zero bias conductance data displayed in Fig.~3, a conductance
suppression is observed which reaches $G(0)/G_\textrm{t}=1/N$ in the zero temperature
limit.

\section{The role of random offset charges}

Thus far, we considered a uniform offset charge $q_\textrm{o}$ applied on all
islands of the CBT. In the absence of externally controlled gate electrodes, the
offset charge is randomized by the electrostatic environment
\cite{RevModPhys.53.497}, which is typically described as an ensemble of
two-level charge fluctuators embedded in tunnel barriers or material interfaces
\cite{doi:10.1063/1.108195,PhysRevB.56.7675,doi:10.1063/1.2949700}. Here, we
assume that any rearrangement of offset charges happens much slower than the
conductance measurement timescale, which is in the order of a second, so that we
can consider a fixed set $\{q_{\textrm{o},i}\}$ for a MCMC calculation run, and
can average the resulting conductance values to acquire the offset
charge-averaged conductance at a given temperature and bias voltage.

\begin{figure}
\centering
\includegraphics[width=0.75\textwidth]{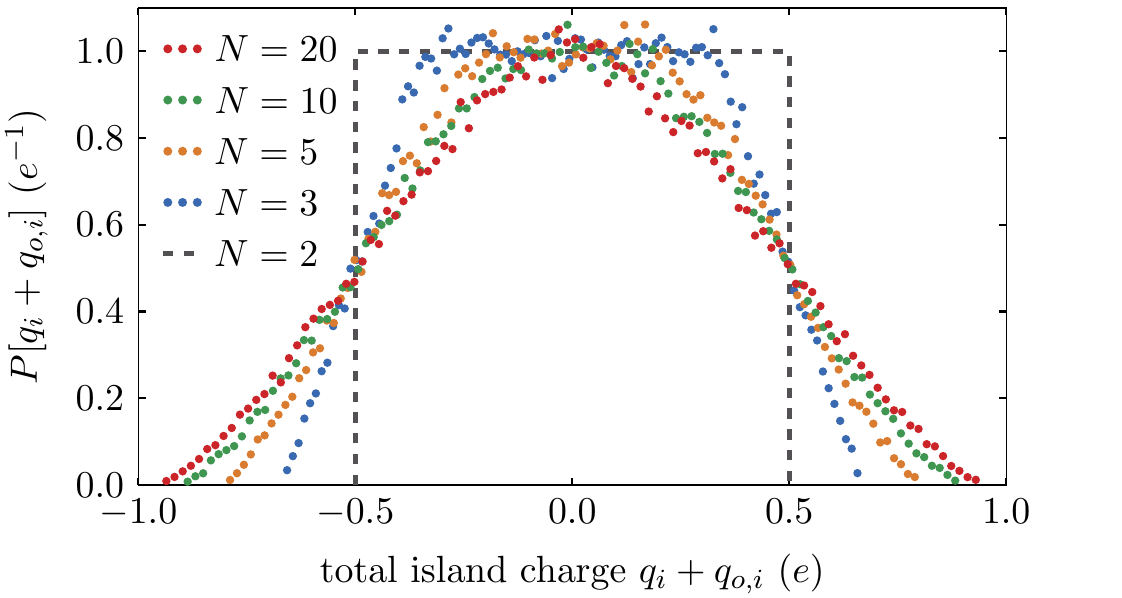}
\caption{The probability density function of the total island charge in a
metastable state as a function of $N$ at zero temperature. The $N=2$ (single island)
case is drawn as a dashed line for comparison.}
\end{figure}

Typical charge trap densities exceed $10^4\,\mu$m$^{-2}$ on Si-SiO$_2$
interfaces \cite{Cheng1977} and in the oxidized surfaces of metal thin layers
\cite{doi:10.1063/1.2949700}, which, together with a typical metallic surfaces
areas exceeding $1\,\mu$m$^2$ of a single island \cite{Yurttaguel2019}, results
in offset charges many orders of magnitude larger than $e$. In a CBT device,
where the islands are tunnel coupled to electrodes, this charge will be reduced
by tunneling until a stable offset charge configuration is reached. In the
single island case ($N=2$) this results in an offset charge uniformly
distributed in the range of $-e/2\ldots e/2$, since any offset charge outside of
this interval would enable a free energy decrease by tunneling of a charge $\pm
e$. While the same range is often assumed in the case of long arrays, the system
can be trapped in a local minimum of the electrostatic energy with tunneling
events towards the global energy minimum being inhibited over realistic
timescales in the low temperature regime.

We demonstrate the presence of such charge configurations by initializing arrays
of different $N$ with a random $q_\textrm{o}$ uniformly distributed with a width of
$\pm100e$, and a stable offset charge configuration is calculated by minimizing
the free energy by a Markov chain of randomly selected tunneling events. The
final charge configuration is recorded when all possible tunneling events out of
this state increase the electrostatic free energy of the array. We show the
resulting probability densities in Fig.~6 with taking the $N=2$ case as a
reference. Notably, the resulting offset charges always fall into the $\pm e$
window, however we find an increasing probability of being outside of $\pm e/2$
with increasing array size. Based on this numerical result, we follow the above
initialization procedure for all MCMC calculations, instead of selecting the
offset charge uniformly between $\pm e/2$ as it was done earlier
\cite{Johansson2000}.

\begin{figure}
\centering
\includegraphics[width=\textwidth]{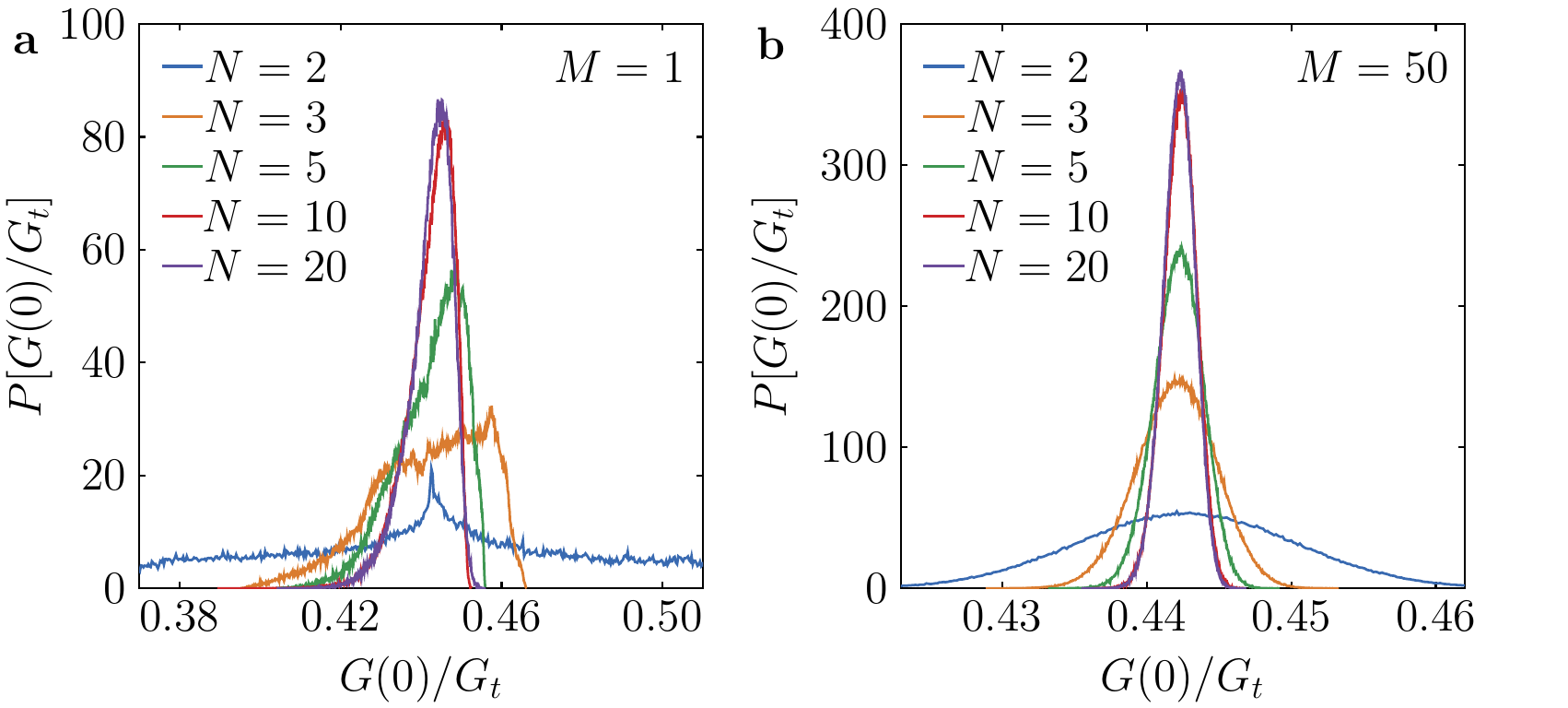}
\caption{The probability density of the conductance of tunnel junction arrays with different
number of junctions in series $N$ for $M=1$ (a) and for $M=50$ (b). The dimensionless temperature $1/u$ values are chosen such that 
the expectation values of the conductance are centered, yielding $0.15$ for $N=2$, $0.18$  for $N=3$,
$0.187$ for $N=5$, $0.189$ for $N=10$ and $0.190$ for $N=20$.}
\end{figure}

We showcase the statistical variations of $G(0)/G_\textrm{t}$ due to the random offset
charges in Fig.~7a for different array lengths. Remarkably, the resulting
distribution narrows with increasing $N$. This is in contrast with the uniform
offset charge case shown in Fig.~3, where the difference between the maximum and
minimum conductance increases as a function of $N$ owing to the strong
exponential suppression of fully blockaded arrays.

We note that the width of the conductance distribution further decreases in the typical CBT
geometry, where several junction series are connected in parallel in order to
increase the device conductance. Assuming an independent offset charge
distribution for all arrays, the standard deviation of the conductance is
expected to decrease with $1/\sqrt{M}$ for an $N\times M$ CBT device. We
demonstrate this effect by plotting the probability density functions of the sum of
$M=50$ randomized conductance values in Fig.~7b.

Both for $M=1$ and $M \gg 1$, we also observe a saturating distribution when
increasing $N$. This defines the long array limit, where the statistical
temperature error of the CBT does not depend on the array length. To simplify
further analysis and comparison with our previously acquired experimental data
\cite{Yurttaguel2019,Sarsby2020}, we use $N=36$ henceforth, which
falls in the long array limit in our temperature regime of interest, $1/u
>0.1$. We also note that several other experiments in the millikelvin regime
were performed with CBTs of a similar or larger array lengths \cite{Casparis2012,Bradley2016,Bradley2017}.

\begin{figure}
\centering
\includegraphics[width=\textwidth]{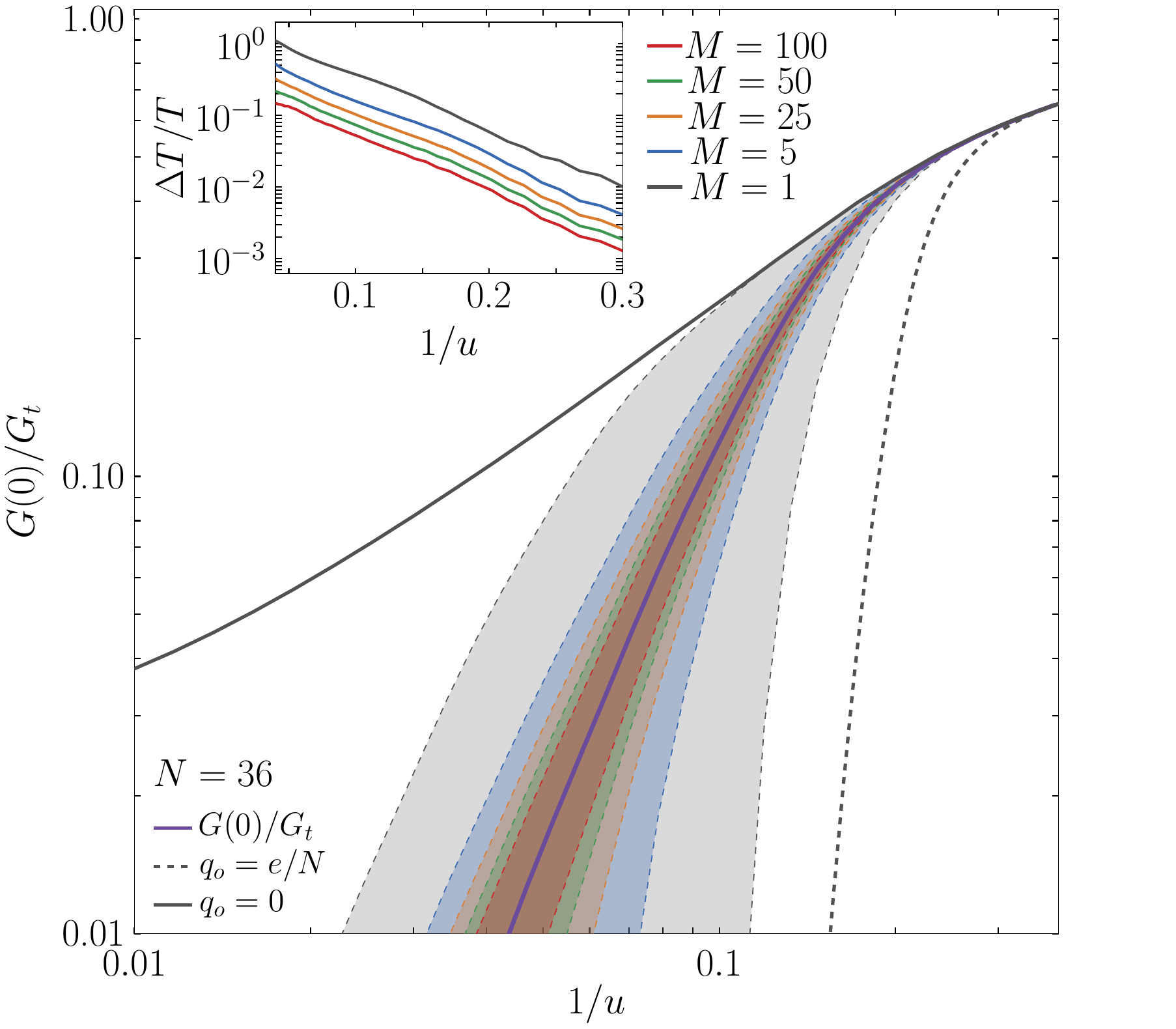}
\caption{Zero bias conductance value and its statistical variation in the low
temperature regime, $u > 1$ for $N=36$. The black solid and dashed lines show the
limiting conductances for charge degeneracy ($q_\textrm{o}=e/N$) and full Coulomb blockade ($q_\textrm{o}=0$),
respectively. The blue solid line is the random offset charge expectation value
with the shaded regions depicting the $3\sigma$ confidence intervals
for $M=1$ (gray), $M=5$ (blue), $M=25$ (orange), $M=50$ (green) and $M=100$ (red). The inset shows the
 relative temperature error $\Delta T/T$ for the same $M$ values.}
\end{figure}

Next, we evaluate the offset-charge averaged conductance as a function of the
temperature in a demonstration of Coulomb blockade thermometry beyond the
universal temperature regime. Here, we plot the expectation
value of $G(0)/G_\textrm{t}$ as a function of the dimensionless temperature $1/u$
(Fig.~8). Remarkably, the confidence intervals corresponding to $\pm3\sigma$
remain much narrower than the full conductance window defined by $q_\textrm{o}=0$ (dashed
line) and $q_\textrm{o}=1/N$ (solid line). Importantly for thermometry applications, we
can evaluate the resulting temperature error $\Delta T$, and find that even for
$M=1$, reliable temperature measurements with $\Delta T/T < 0.05$ are possible
if $1/u > 0.2$, in contrast with the $1/u > 0.4$ limit calculated
earlier for $N=2$ \cite{Feshchenko2013}. Furthermore, CBT devices with several junction series in
parallel are even more suitable in the low temperature regime: a realistic CBT
with $M=100$ reaches the same temperature error at $1/u = 0.1$
demonstrating the possibility of precision low temperature electron thermometry by
using scalable fabrication of large arrays.

We display the calculated finite bias conductance curves of the same device in
Fig.~9, which is the low temperature extension of the results presented in
Fig.~4. Remarkably, the full conductance traces remain sensitive to the
temperature even when $1/u \ll 1$, which allows for a primary calibration of
the CBT in the low temperature regime, despite the absence of offset-charge
control on the individual islands.

\begin{figure}
\centering
\includegraphics[width=\textwidth]{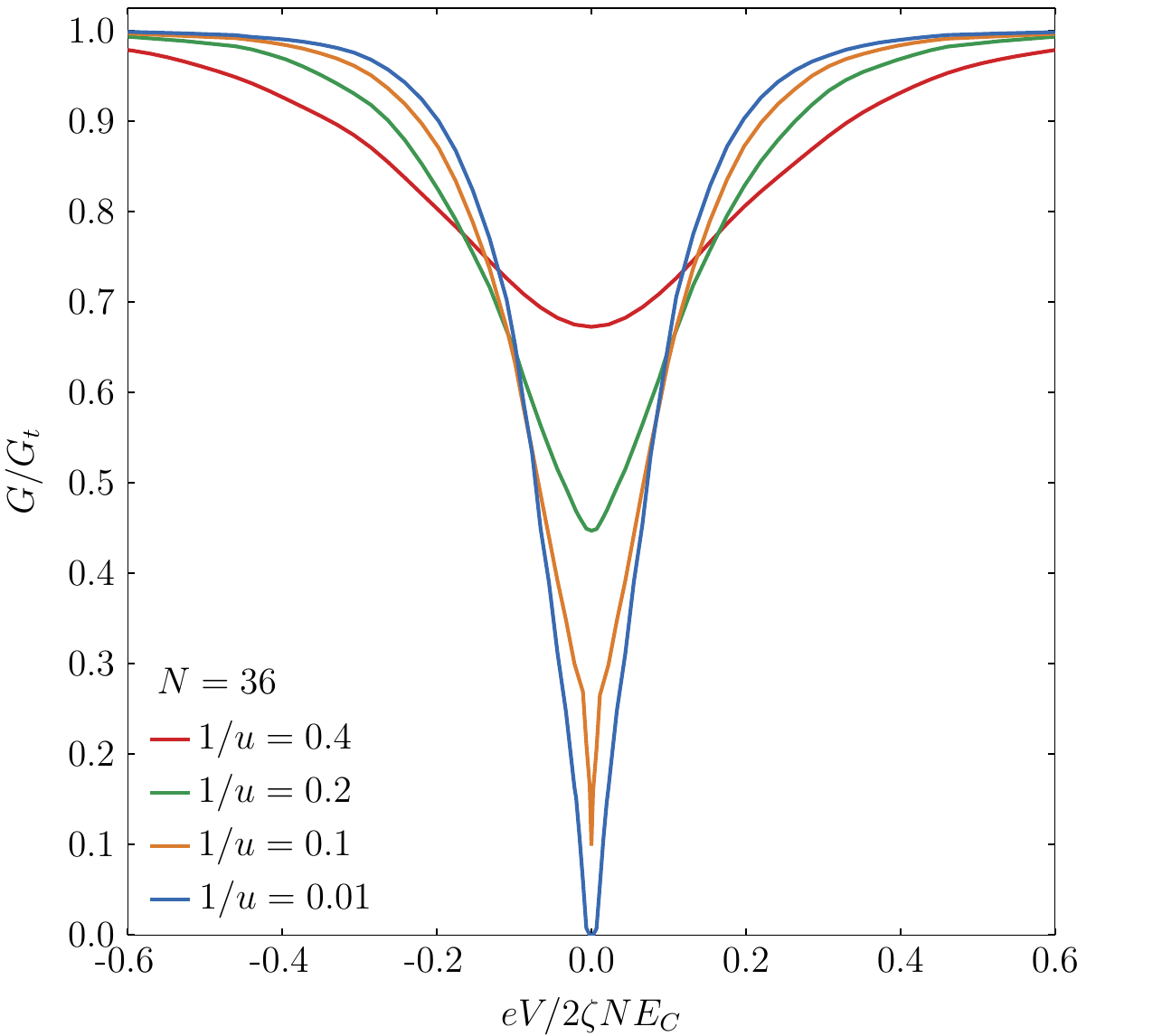}
\caption{The offset charge-averaged normalized differential conductance curves of a $N=36$ device as
the function of the dimensionless bias voltage in the low temperature regime
where $u > 1$ calculated by the MCMC model. Note that the voltage scaling is the same as in Fig.~4.}
\end{figure}

\section{Disorder in the tunnel junction parameters}

Previously we argued that the effect of random offset charges can be mitigated
by using devices with large $N$ and $M$. However, for realistic device modeling,
we also have to consider the non-uniformity in the tunnel junction resistance $R_i$
and island coupling capacitance $C_i$ values. These variations can be due to
lithographic inaccuracies as well as thin film thickness variations. Previous
studies addressed this error source assuming that the junction area inhomogenity
of \emph{in-situ} created AlO$_x$ tunnel barriers dominates \cite{hahtela2013investigation,doi:10.1063/1.115090}, equivalent to a
constant $R_iC_i$. However, recent studies demonstrated the spatial variations in the oxide layer thickness
\cite{Zeng_2015} and new fabrication techniques utilizing \emph{ex-situ} via
tunnel junctions were developed \cite{Yurttaguel2019,prunnilaexsitu}, requiring a separate
evaluation of the effects of the capacitance and resistance variations on the CBT accuracy.

\begin{figure}
\centering
\includegraphics[width=\textwidth]{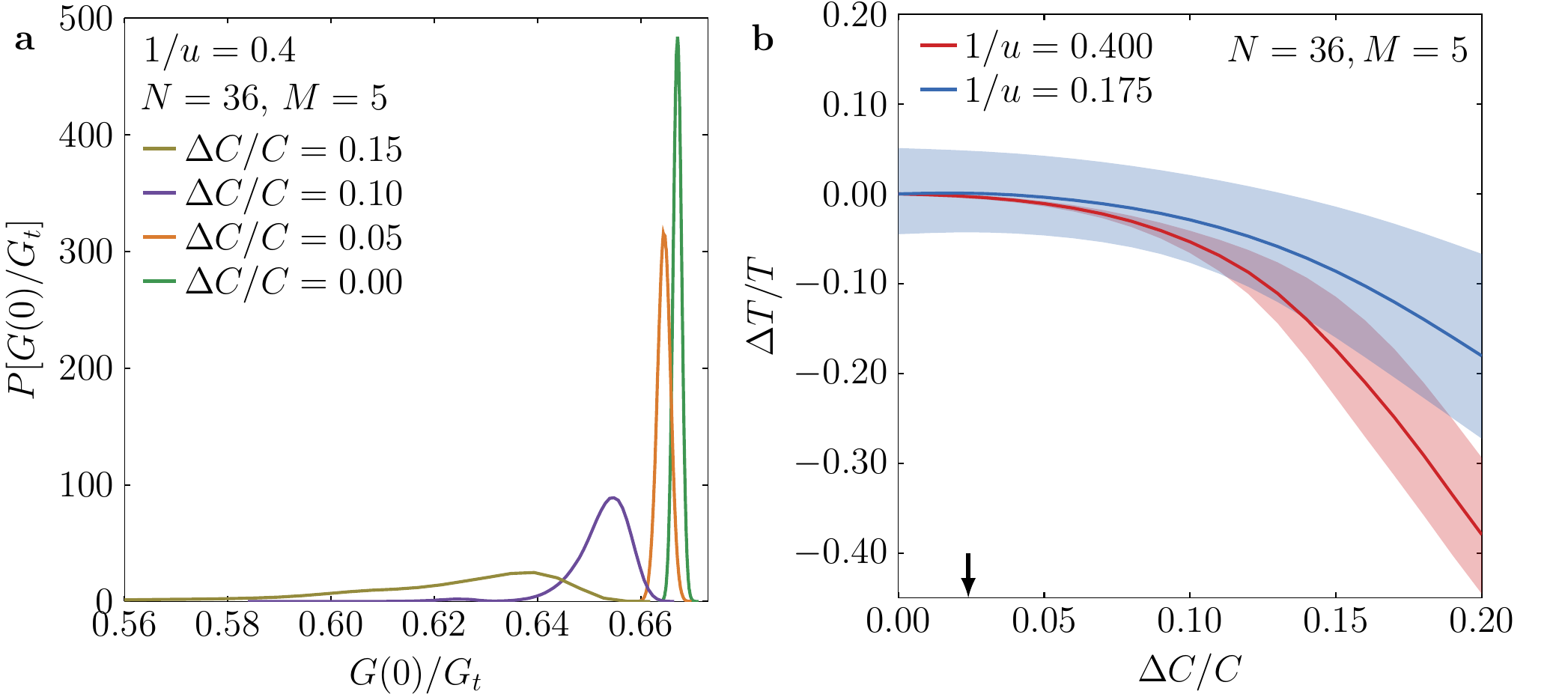}
\caption{Relative temperature error caused by the variation of the junction
capacitance, $C$. (a) The conductance distributions for different capacitance
variations at $1/u=0.4$. The capacitances were assumed to be distributed
uniformly around $C$ in an interval of $\Delta C$. (b) The systematic change in
the expectation value is depicted by the solid lines at $1/u=0.4$ (red) and
$1/u=0.175$ (blue). The shaded regions denote the $3\sigma$ confidence
intervals. The calculations were performed using $N=36$ and $M=5$ in both
panels. The arrow shows the estimated $\Delta C$ for our experimental data, see the text.}
\end{figure}

We first discuss the effect of the capacitance disorder by calculating a
conductance histogram of an $N=36$ and $M=5$ device with a uniform distribution
of junction capacitances, $C\pm\Delta C/2$ (Fig.~10a). Here, $\Delta C/C=0$
corresponds to a homogeneous array, where the conductance distribution is
governed by the random offset charges. Increasing $\Delta C/C$ results in a
broadening of the conductance histogram as well as a quadratic drop in the
expectation value of $G(0)/G_\textrm{t}$, which translates to a decrease in the
inferred temperature (see Fig.~8). We note that we kept $1/u$ constant for
different $\Delta C$ values, using the expectation value $C$ in Eq.~(5).
We show in Fig.~10b that the impact of $\Delta C/C$ on thermometry depends on the
temperature by displaying the relative temperature error $\Delta T/T$ defined as
the $3\sigma$ confidence intervals as a function of $\Delta C/C$ at different
$1/u$ values. Our results indicate that in the low temperature regime, $1/u <
0.4$ and realistic capacitance variations of a few percent, the measurement
error is dominated by the randomness of the offset charges.

\begin{figure}
\centering
\includegraphics[width=\textwidth]{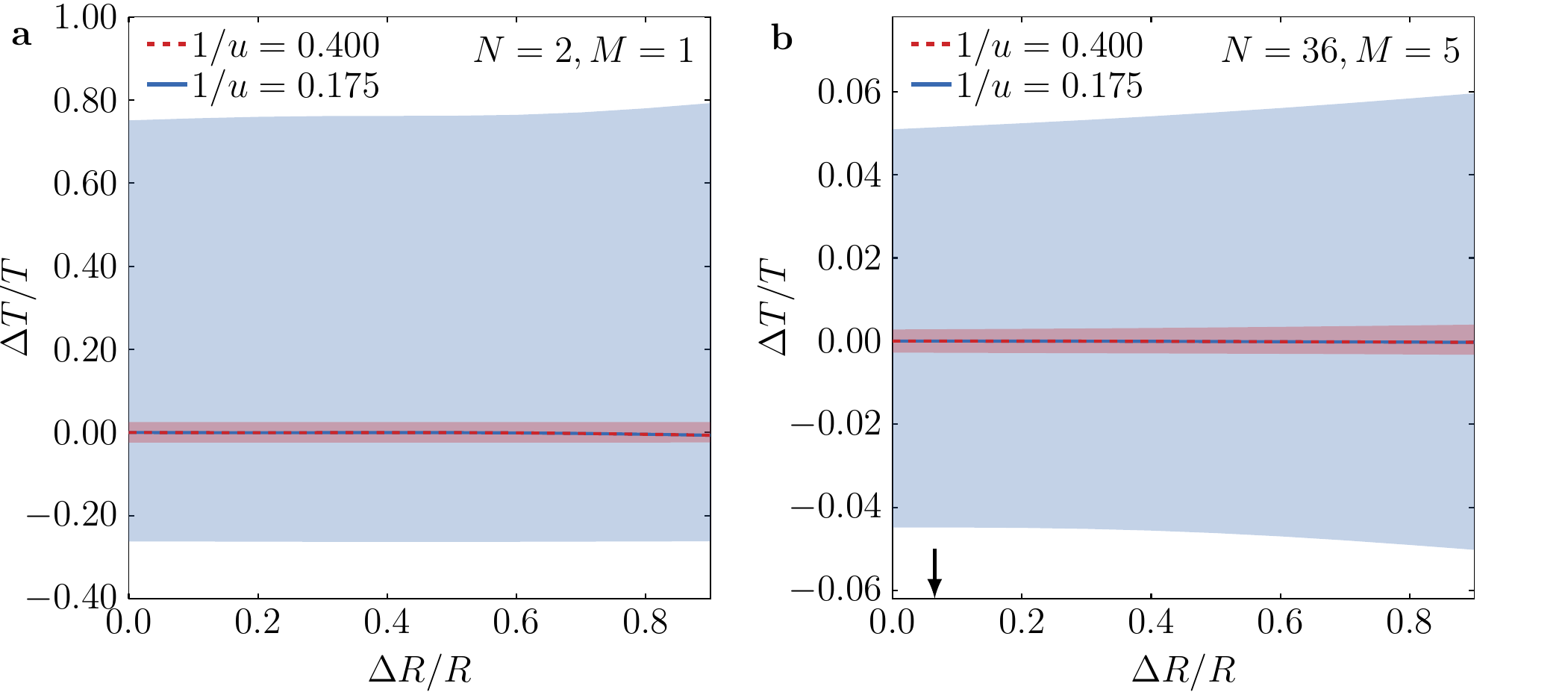}
\caption{Relative temperature error caused by the variation of the tunnel
junction resistance, $R$. The systematic change in the expectation value is
depicted by the solid lines at $1/u=0.4$ (red) and $1/u=0.175$ (blue) for the
single island case ($N=2$ and $M=1$, panel a) and for the long array, $N=36$,
$M=5$ case (panel b). Note the difference in the vertical scale between the two
panels. The shaded regions denote the $3\sigma$ confidence intervals. The tunnel
junction resistances are distributed uniformly around $R$ in an interval of
$\Delta R$. The arrow shows the estimated $\Delta R$ for our experimental data,
see the text.}
\end{figure}

Next, we evaluate the case of variations in the individual tunnel junction
resistances, and perform the MCMC simulations with a uniform distribution of $R$
with a width of $\Delta R$ (Fig.~11). The resistance variations cause a weak
quadratic decrease in the expectation value of the temperature, however even the
extreme case of $\Delta R/R \lesssim 1$ yields a smaller systematic and
statistical error than the capacitance disorder in the $\Delta C/C \sim 0.1$
regime. When comparing the $N=2$ and $M=1$ (Fig.~11a) single island CBT with the
$N=36$ and $M=5$ case (Fig.~11b), we find a marked reduction in the statistical
temperature error for the long array, further attesting to the importance of
large array devices for low-temperature thermometry.

\section{Comparison with experimental data}

\begin{figure}\centering
\includegraphics[width=\textwidth]{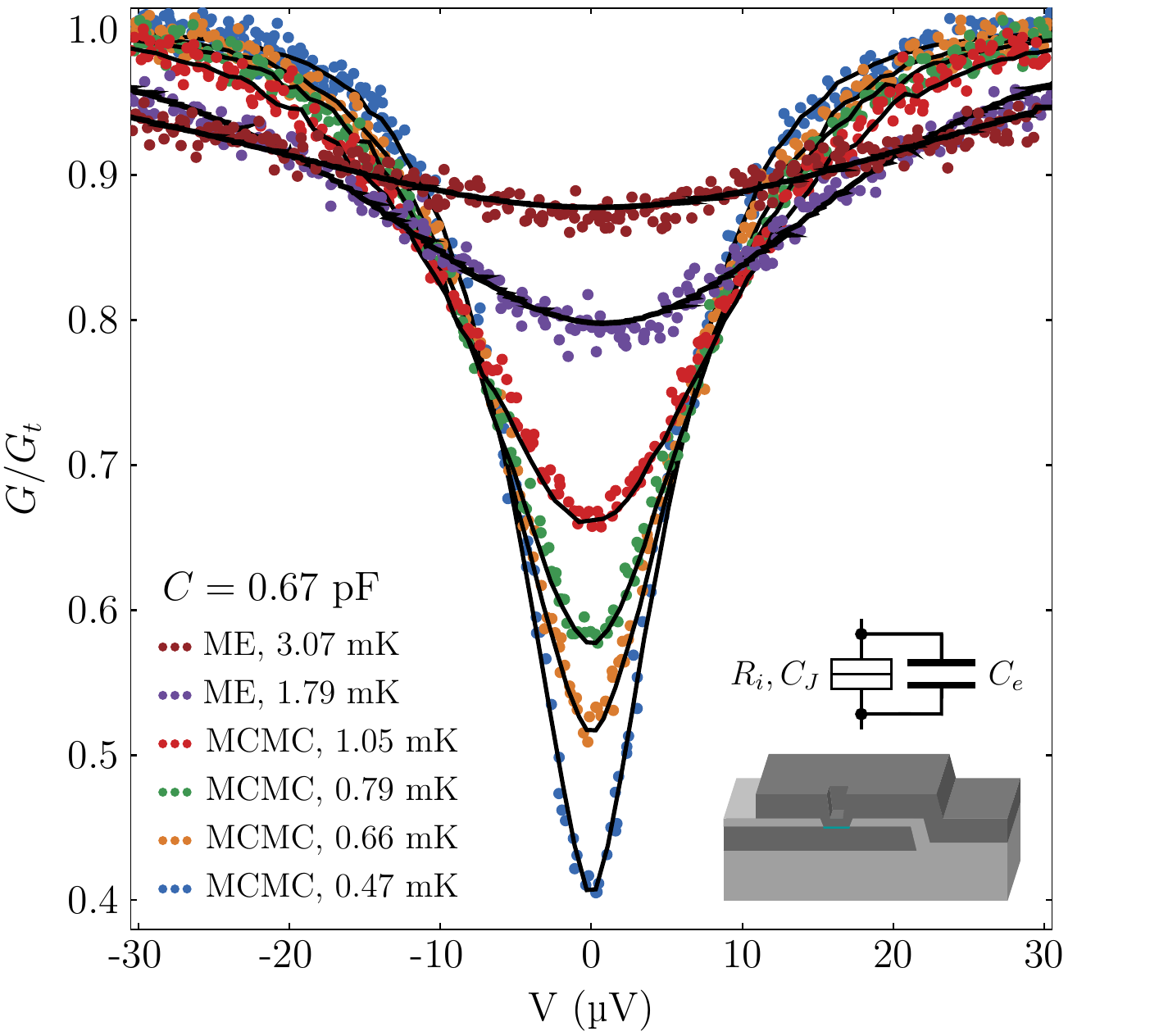}
\caption{Comparison of ultralow-temperature experimental data and theoretical
curves using a device with $N=36$. The island capacitance of $C=670\pm2\,$fF is
obtained from fitting against the master equation (ME) model in the universal
regime, see the uppermost two curves. The MCMC fit curves at lower four
temperature values are acquired by fitting the conductance at zero bias, and
then simulating the rest of the curve with no additional parameters. In this
regime, the ME model does not describe the conductance curves. For all sets,
solid lines display the calculated curves and dots represent the experimental
data taken at a fixed magnetic field of $50\,$mT. The inset shows the schematic
cross-section of the measured CBT featuring via tunnel junctions between
neighboring islands (dark grey) through the dielectric layer (light grey)
yielding $C=C_\textrm{J}+C_\textrm{e}$, where the capacitance contribution
$C_\textrm{e}$ from the overlapping islands dominates.}
\end{figure}

We now demonstrate the correspondence of the MCMC simulations with experimental
data taken in the sub-millikelvin electron temperature regime. We fabricated our
Coulomb blockade thermometers with \emph{ex-situ} Al/AlO$_x$/Al via tunnel
junctions connecting adjacent Al islands separated by sputtered SiO$_x$
interlayer dielectric \cite{Yurttaguel2019,prunnilaexsitu}. This process, in
contrast to the frequently used Dolan bridge technique
\cite{doi:10.1063/1.89690}, enables an independent tuning of the tunnel junction
area and the capacitive coupling between the islands. Crucially for the
applicability of our numerical model, our device fulfills the $C_\textrm{o} \ll
C$ requirement with a large fraction of the island areas overlapping (see the
inset of Fig.~12). Our device features $N=36$ tunnel junctions in series and
$M=5$ rows in parallel.
Electron cooling was performed by on- and off-chip nuclear refrigeration
\cite{Palma2017}, using large volume electrodeposited indium fins. When the
nuclear demagnetization was carried out on the device, we observed \cite{Sarsby2020} that the measured
electron temperature scales with the effective magnetic field $\sqrt{B^2 +
b_\textrm{i}^2}$, where $B$ is the applied magnetic field and $b_\textrm{i}= 295\,$mT is the
internal magnetic field corresponding to the quadrupolar crystal field in
indium, consistent with prior literature data \cite{Tang1}. We found that the distortion of
$G(V)$ because of self-heating \cite{doi:10.1063/1.353092} was negligible. The
details of the device fabrication and measurement setup are published elsewhere
\cite{Yurttaguel2019,Sarsby2020}.

First, we measured our CBT in the universal regime, where the ME model provides
a reliable calibration of the island capacitance, $C=670\pm2\,$fF, which we will
use as a fixed parameter for the rest of the analysis. In the universal regime,
where $G(0)/G_\textrm{t}>0.8$, we can get reliable fits using the ME model (see the two
uppermost curves in Fig.~12).
Below this threshold, however, we first fit the temperature comparing the zero
bias conductance with the MCMC model prediction (see Fig.~8), and then confirm
the validity of the fit in the full voltage bias range (four lower curves in
Fig.~12). With this procedure, we get an excellent agreement between our
numerical simulations and experimental data down to $T=470\,\mu$K corresponding
to $1/u=0.175$, where we estimate a relative temperature error of $\Delta T/T =
0.049$ (see Fig.~8).
This result demonstrates that accurate Coulomb blockade thermometry is possible
outside of the commonly considered universal regime.

Finally, we discuss the temperature error associated with the variations in the
tunnel junction parameters and compare the resulting temperature error with the
previous result. Based on the maximum alignment error of $200\,$nm of the laser
lithography used to fabricate our device, we infer a relative error in overlap
and in the island capacitance $\Delta C/C=0.024$ (see vertical
arrow in Fig.~10b), which in our temperature of interest contributes with less
than $0.01$ to the relative temperature error. We characterized our tunnel
junctions to exhibit an areal resistance
$12.8\pm0.8\,\textrm{k}\Omega\mu\textrm{m}^2$ \cite{Yurttaguel2019},
corresponding to $\Delta R/R=0.063$ (see vertical arrow in Fig.~11b), which has
a negligible effect on the temperature error. Our analysis attests to the
insensitivity of CBTs to realistic fabrication inaccuracies in the low
temperature regime, in agreement with similar numerical studies in the universal
regime \cite{Pekola1994,hahtela2013investigation,Meschke2016,Hirvi1996}.

\section{Corrections with a finite gating capacitance}

We now turn to the case where the island gating capacitance $C_{\textrm{o},i}$ is finite,
but still much less than the $C_i$ capacitances between the islands (Fig.~1). First,
we establish the validity of this limit by estimating the self-capacitance of
the islands in our experiments \cite{Yurttaguel2019,Sarsby2020}, which is
assumed to be dominated by the indium cooling fins \cite{Yurttaguel2019} with
linear dimensions of $25\times 50 \times 140\,\mu$m$^3$. To provide an estimate, we use the
self-capacitance $C_\textrm{o}=4\pi\varepsilon_0 d\cdot 0.66$ of a cube
\cite{doi:10.1063/1.1664031} with $d=50\,\mu$m side yielding $C_\textrm{o}\approx3.7\,$fF
and $C_\textrm{o}/C\approx5.5\times 10^{-3}$.

\begin{figure}\centering
\includegraphics[width=\textwidth]{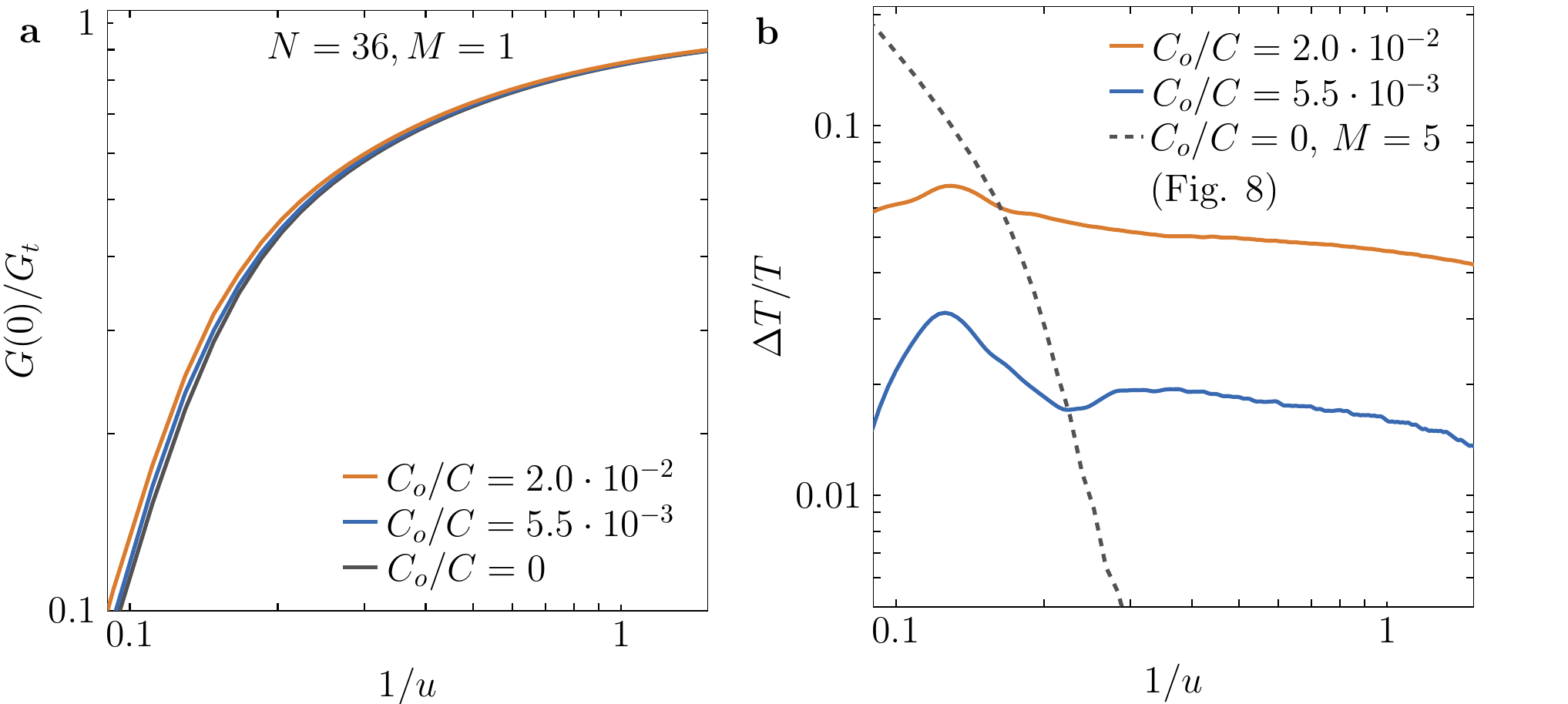}
\caption{The effect of a finite gate capacitance on the temperature calibration
for different $C_\textrm{o}/C<< 1$ values with a constant $C_\Sigma$ and $N=36$.
(a) The normalized zero bias conductance as a function of the dimensionless
temperature $1/u$. (b) The relative systematic temperature error $\Delta T/T$
when using the $C_\textrm{o}=0$ model (solid lines) compared with the statistical error
due to the random offset charges for $M=5$ (dashed line), repeated from the
inset of Fig.~8. We use the offset-charge averaged model for both panels.}
\end{figure}

We use the numerical model discussed in Section 3 using the capacitance matrix
$\mathbf{C}$ (see Eq.~[2]) with a modified $C_{\Sigma i}=C_i + C_{i+1} +
C_{\textrm{o},i}$. As we demonstrated in Fig.~10 and Fig.~11, the leading term in
statistical variations is caused by the random offset charges, therefore we
evaluate the offset-charge averaged zero bias conductance curves as a function
of temperature for finite $C_\textrm{o}/C$ ratios while keeping $C_{\Sigma}$ constant with no
disorder in the capacitance values and tunnel junction resistances. Similarly to
our other calculations, we set $N=36$ to enable a direct comparison with our experiments.

We summarize our calculations in Fig.~13, where we display the temperature
dependence of the normalized zero bias conductance (Fig.~13a) for a $C_\textrm{o}/C$
ratio of $5.5\times 10^{-3}$ and $2\times 10^{-2}$, which we compare to the
$C_\textrm{o}/C=0$ model also shown in Fig.~8. We recover the high temperature, universal
regime, where the curves merge, however we find an increasing systematic
deviation when lowering the temperature to the the intermediate
\cite{Feshchenko2013} and to the low temperature limit. In Fig.~13b, we
characterize the resulting relative temperature error resulting from using the
$C_\textrm{o}=0$ model in case of these two finite $C_\textrm{o}/C$ ratios. We note that $\Delta
T/T>0$, thus neglecting $C_\textrm{o}$ yields an overestimate of the device temperature.
Our numerical results show that this error is of the order of $1\%$ in our
temperature regime of interest, and it increases with the $C_\textrm{o}/C$ ratio of the
device, demonstrating the importance of the specific CBT device design for ulta-low
temperature thermometry.

We conclude this analysis by comparing this systematic deviation to the
statistical uncertainty caused by the random offset charges for $M=5$, relevant
for our device. We find that in the low temperature limit $1/u<0.2$, the
statistical uncertainty dominates. This is in contrast to the intermediate
regime $1/u\lesssim 1$, where the finite $C_\textrm{o}/C$ ratio plays a more important
role. Our numerical results demonstrate that CBT devices with large area
overlaps between neighboring islands \cite{Bradley2016,Yurttaguel2019} are
preferred in the intermediate regime to devices with shadow-evaporated tunnel
junctions \cite{Meschke2016} and large surface area cooling fins
\cite{Palma2017} featuring higher $C_\textrm{o}/C$ ratios.

\section{Conclusions}

Using Markov-chain Monte Carlo simulations, we demonstrated that Coulomb
blockade thermometry can be extended beyond the universal regime. Our results
show that long arrays with many islands in series benefit from the narrowing
total conductance distribution even in the case of random offset charges, and
the statistical temperature errors are further reduced by adding several
junctions series in parallel, which is a typical geometry for practical CBTs.
These contributions lead up to a factor of four increase in the usable
temperature range of a CBT with a realistic number of tunnel junctions,
impacting precision thermometry applications. We demonstrated the
applicability of the numerical results to experimental data taken in the
sub-millikelvin temperature regime. These temperatures fall outside of the
universal regime for a CBT featuring a charging energy exceeding $1\,$mK, which
is necessary to perform a primary calibration at electron temperatures in the
range of $10\ldots100\,$mK attainable in commercial dilution refrigerators. Finally, we
showed that at sub-millikelvin temperatures, device designs yielding a low
gate capacitance ratio $C_\textrm{o}/C\lesssim 10^{-3}$ are necessary to suppress
systematic temperature errors stemming from a finite $C_\textrm{o}$.

\section*{Data availability}

Raw datasets and computer code are available at the Zenodo repository \cite{rawdata}.

\begin{acknowledgements}
The authors thank J.~Pekola for fruitful discussions. This work was supported by
the Netherlands Organization for Scientific Research (NWO) and by the European
Research Council under the European Union's Horizon 2020 research and innovation
programme, grant number 804988.
\end{acknowledgements}
 
\section*{Conflict of interest}
The authors declare no conflict of interest. 
 
\bibliographystyle{spphys}
\bibliography{references}

\end{document}